# Correction of aberrations via polarization in single layer metalenses

*Augusto Martins,[†,⊥] Kezheng Li,[‡] Guilherme S. Arruda,[†] Donato Conteduca,[‡] Haowen Liang,[§] Juntao Li,[§] Ben-Hur V. Borges,[†] Thomas F. Krauss[‡] and Emiliano R. Martins[†*]*

[†]São Carlos School of Engineering, Department of Electrical and Computer Engineering, University of São Paulo, Brazil, 13566-590

[‡]University of York, School of Physics, Engineering and Technology, York, UK, YO10 5DD

[§]State Key Laboratory of Optoelectronic Materials and Technologies, School of Physics, Sun Yat-sen University, Guangzhou, China, 510275

**ABSTRACT**

The correction of multiple aberrations in an optical system requires different optical elements, which increases its cost and complexity. Metasurfaces hold great promise to providing new functionality for miniaturized and low-cost optical systems. A key advantage over their bulk counterparts is the metasurface's ability to respond to the polarization of light, which adds a new degree of freedom to the optical design. Here, we show that polarization control enables a form-birefringent metalens to correct for both spherical and off-axis aberrations using a single element only, which is not possible with bulk optics. The metalens encodes two phase profiles onto the same surface, thus allowing switching from high resolution to wide field of view operation. Such ability to obtain both high resolution and wide field of view in a single layer is an important step towards integration of miniaturized optical systems, which may find many applications, e.g., in microscopy and endoscopy.

**Introduction**

A well-established paradigm in optics asserts that it is impossible to correct multiple aberrations with a single optical element only. Consequently, high quality optical systems require a combination of optical elements, thus increasing their complexity and cost. Many of these traditional limitations in optics, however, have been challenged by the recent emergence of metasurfaces.

Metasurfaces have been attracting significant attention due to their potential for achieving unique optical functionalities in miniaturized systems. A metasurface locally manipulates the properties of light at the level of the so-called meta-atoms, which are nanoresonators that interact strongly with light. By engineering the geometry of the meta-atoms, the phase, amplitude, polarization, and even the phase dispersion of light can be manipulated at the subwavelength scale[1-3]. The ability to use polarization as a degree of freedom is of particular interest because it adds a functionality that is virtually absent from conventional bulk optics. For instance, this functionality has led to the demonstration of a compact full-Stokes camera that can decompose an image into all four Stokes vector components[4]. Multifunctional polarization-based metalenses[5-8], birefringent holograms[9,10], orbital angular momentum generation[11] and detection[12] are a few more examples that have used this unique feature of metasurfaces.

Here we show that metasurfaces can overcome bulk optics impossibility of correcting for multiple aberrations with a single element. We demonstrate that a single layer metasurface with birefringent meta-atoms can correct both spherical and off-axis aberrations.

It is well-known that hyperbolic lenses are free from spherical aberrations[13,14], but these lenses are not employed frequently in bulk optics because of their demanding fabrication requirements. In contrast to bulk optics, metalenses can easily impose any phase profile, including hyperbolic, which has enabled the demonstration of diffraction-limited focusing[15]. A major drawback of hyperbolic metalenses, however, is their very limited field of view (FOV) due to off-axis aberrations[13, 16]. The FOV can be improved by means of a double layer, or doublet, metasurface[17, 18], but this strategy loses one of the metalenses key advantages: their simplicity.

Recently, it has been shown that a quadratic phase profile is capable of focusing light from any angle of incidence without distortion[16, 19-21]. This feature has led to the demonstration of metalenses free of off-axis aberration, thus featuring a very wide FOV. Quadratic metalenses, however, suffer from spherical aberration, which limits their resolution to about $\sim 2\lambda_0$[3, 16, 20]. Thus, on the one hand, hyperbolic metalenses are free from spherical aberrations (leading to high resolution) but suffer from off-axis aberrations (leading to narrow FOV). On the other hand, quadratic metalenses suffer from spherical aberrations (leading to a lower resolution) but are free from off-axis aberration (leading to wide FOV).

Here, we combine the advantages of high resolution and wide field of view by encoding both hyperbolic and quadratic phase profiles in a single metalens using polarization multiplexing. As a proof of concept, we demonstrate a metalens based on elliptical nanoposts that allows independent phase control for two orthogonal linear polarization states[10]. We also demonstrate wide field of view and high-resolution imaging, respectively, by controlling the polarization state. We characterize the metalens via both the point spread function (PSF) and the imaging properties for the respective polarization states. PSF measurements show that the FOV is approximately 54° when operating as a quadratic metalens and less than 1° when operating as a hyperbolic metalens, while the hyperbolic lens can achieve near-diffraction limited resolution. A potential application example is to use a single-element birefringent metalens to build a compact and low-cost integrated microscope with two operation modes, a capability that might be suitable for endoscopy[22]. Such a microscope would allow the user to begin scanning a large area with the quadratic profile and then rotate the polarizer to explore a smaller area within the first to obtain an image with better resolution.

**Metalens concept and design**

To achieve a wide field of view and high resolution with a single element, the birefringent metalens encodes a quadratic phase profile (Equation 1) for the x-polarized incident light, and a hyperbolic phase profile (Equation 2) for the y-polarized incident light (see Fig. 1 (b)). In these equations, $k_0$ is the free space wavenumber, $f_q$ and $f_h$ are the focal lengths, $n_{ext}$ is the focusing medium refractive index, and $r$ is the in-plane radial distance in the plane of the metalens. The quadratic profile (Equation (1)) can image over a wide field of view[16], but spherical aberration limits its resolution to $\sim 2\lambda_0$[3, 16, 19, 20]. The hyperbolic profile (Equation 2) is free of spherical aberration, so it can image with high resolution but only over a small field of view (FOV)[13].

$$\phi_q(r) = \frac{-k_0 r^2}{2 f_q} n_{ext} \quad (1)$$

$$\phi_h(r) = -k_0 n_{ext} \left( \sqrt{f_h^2 + r^2} - f_h \right) \quad (2)$$

The birefringent metalens consists of 230 nm tall crystalline silicon (*c*-Si) elliptical nanoposts on a sapphire substrate designed to operate at the wavelength of 532 nm[10] (Fig. 1b). The high index of c-Si (n = 4.14+0.032i [23]) causes a high field confinement within the nanoposts[24]. We assume each nanopost as a truncated waveguide supporting only the propagation of the fundamental mode ($HE_{11}$)[25]. Consequently, the phase shift of the transmitted light depends mainly on the effective index of this mode, which, in turn, depends on the diameter of the post. The elliptical cross section breaks both the waveguide rotation symmetry and the $HE_{11}$ mode degeneracy along each semi-axis of the ellipsis[26]. In other words, the elliptical posts present form birefringence with the slow and fast axes aligned with the ellipsis semi-axes. By tuning the length of the ellipsis semi-axes (alongside their ellipticity), it is possible to obtain independent full phase control [0-2π rad] for light polarized along the slow and fast axes of the nanoposts[10, 25]. According to Malus' law[27], to maximize transmission and avoid cross-polarization terms, we fixed the ellipsis axis for all posts to the laboratory horizontal ($x$) and vertical ($y$) directions.

We calculate the phase and transmission of the elliptical nanoposts with the Rigorous Coupled Wave Analysis method (RCWA)[28, 29] with plane wave excitation ($\lambda_0$ = 532 nm) from the sapphire substrate ($n_s$ = 1.77[30]) at normal incidence. The calculation is made for a square lattice and a unit cell of fixed size (210 nm). The semiaxes $D_x$ and $D_y$ ( Fig. 1 (a)) are swept in the range [40 nm, 210 nm]. The zeroth order transmittance and relative phase maps are shown in Fig. S1 of the Supplementary Information (SI). Using a similar procedure as in [10], we identified a set of 36 structures on the maps that can give six phase levels for each linear polarization. A list with all structures is available in Fig. S2 of the SI.

The quadratic and hyperbolic phase profiles are quantized with six phase levels and encoded on the x and y polarizations of the birefringent metasurface following Fig. S2. The structures were fabricated using high-resolution electron-beam lithography (see Methods). Figures 1(c) and (d) show micrographs of the fabricated birefringent metasurface.

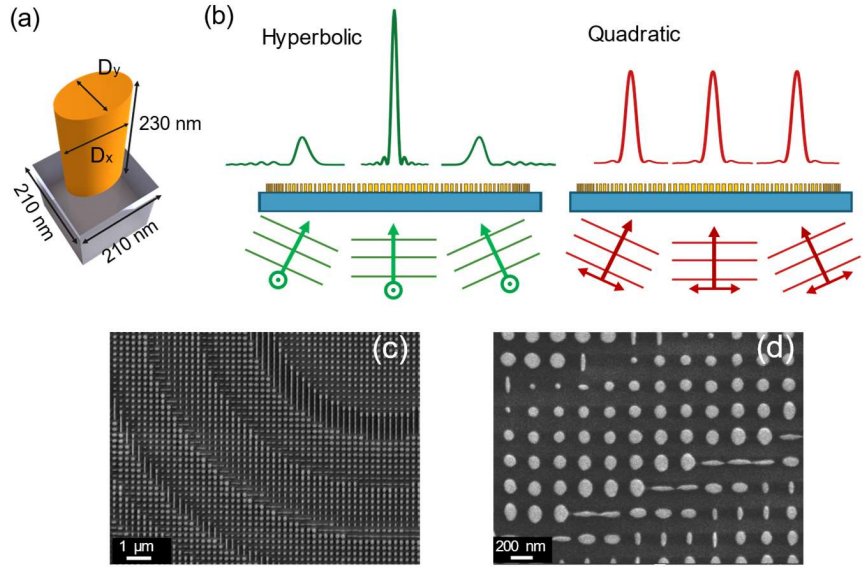

**Figure 1**. (a) Unit cell representation and (b) illustration of the birefringent metalens operation. The form birefringence of the meta-atoms allows the independently encoding of two phase profiles in the metasurface. The light must be polarized along the axes of the nanoposts to access the respective phase profiles. The left drawing in (b) (green rays and arrows; transverse linear polarized light) illustrates how to access the hyperbolic phase to obtain a high resolution at near normal incidence but with strong off-axis aberration. The right drawing in (b) (red rays and arrows; in-plane polarized light) illustrates how to access the quadratic profile to obtain a low resolution with a wide field of view and a broader point spread function. (c) and (d) show SEM micrographs of the fabricated birefringent metasurfaces at different magnifications.

**Birefringent imaging**

To demonstrate the flexibility of the metalens, we test its imaging properties as an objective lens for two different systems: a telescope setup, with the object placed far from the metalens and its image formed near the focal plane; and a microscope, with the object near the focal plane (see Methods for more details on the optical systems). We start with the telescope configuration using as object the USAF 1964 chart placed 7.3 mm away from the metalens. The images obtained with x-polarized (quadratic) and y-polarized (hyperbolic) light are shown in Figures 2 (a) and (b), respectively. The birefringent metalens has a focal length of 0.66 mm for both the hyperbolic and quadratic phase profiles (an example with different focal lengths is shown in the SI, section S3) and a diameter of ~1.5 mm (numerical aperture, NA = 0.75). For this configuration, the image forms very close to the focal plane with a magnification of approximately -1/10. When assessing the image, we notice that the long bars on the edges (top inset of Fig. 2 (a)) have dimensions of 2.47 μm × 12.35 μm in the image plane. They are well-defined for x-polarized light (indicating the wide FOV for this polarization, Fig 2(a)) but appear blurred for y-polarized light (hyperbolic, top inset of Fig. 2(b)). Conversely, at the center of the field of view, most of the shorter bars appear blurred for x-polarized light (quadratic) but are well-resolved for y-polarized light (hyperbolic), thus indicating a high resolution for this polarization. The widths of the respective rectangles are 0.55 μm, 0.49 μm, and 0.44 μm. The higher resolution is better appreciated with a 2× digital zoom, shown in the bottom insets in Figs. 2(a) and (b). These insets clearly show that the metalens can resolve the small rectangles only when illuminated with y-polarized light (hyperbolic phase). Note that the hyperbolic profile can form images with discernable features for dimensions as small as 440 nm, which is close to the diffraction limit of ~384 nm (NA=0.75, $\lambda_0 = 532\ nm$ )[16]. Considering a possible application, we note that optical imaging typically requires two steps. First, we use a wide field of view lens to find the target, and then, we zoom in on the target with a high resolution lens to image it in detail. In bulk optics, these two steps require two different optical systems, e.g., two separate objectives mechanically swapped on a turret. Our experiment shows that a single layer birefringent metalens assisted by a polarizer can successfully replace these two optical systems. For instance, we envisage that this additional degree of freedom provided by our design might be suitable for improving metalens-based endoscopes [22], for which it is easy to adjust the polarization remotely, but not the imaging optics.

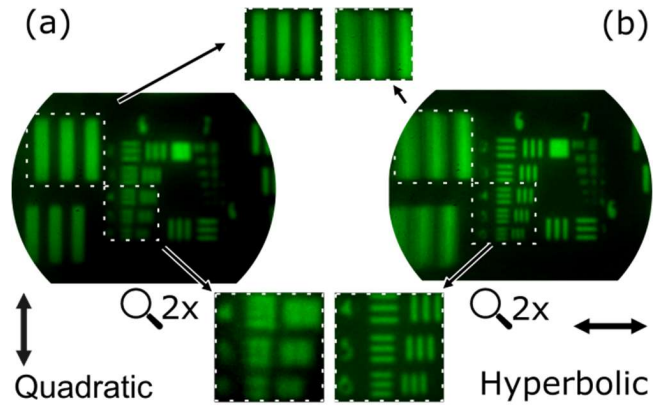

**Figure 2.** (a) (x-polarized light, quadratic profile) and (b) (y-polarized light, hyperbolic profile) showing the USAF 1951 chart images obtained with the telescope setup using the metalens as an objective lens. The metalens focal length is 0.66 mm, and the chart is 7.3 mm away from the metalens, rendering a magnification of ~-1/10. The longer bars shown in the top insets have dimensions of 2.47 μm × 12.35 μm in the imaging plane, while the smaller bars shown in the bottom insets are 0.62 μm, 0.55 μm, 0.49 μm and 0.44 μm in size. The bottom insets were digitally magnified by a factor x2 to emphasize the difference in resolution.

The second imaging system is a microscope, where we place the objects close to the focal length of the metalens and image an array of c-Si cylinders. The cylinders are 230 nm tall, have a diameter of 1 μm, and are arranged in a simple square lattice with a period of 3 μm. The cylinder array images obtained with the quadratic and hyperbolic phase profiles are shown in Figures 3(a) and (b), respectively. The posts imaged by the quadratic x-polarization are uniformly blurred across the image (see Fig. 3(a)), which is a characteristic of the wide field of view, but with a lower resolution. The image obtained with the hyperbolic profile and y-polarized light, however, shows sharply defined cylinders at the centre – indicating its high resolution – but strongly blurred and distorted at the edge of the image – indicating a poor FOV (compare Figs. 3(a) and (b)). Due to the wide field of view of the quadratic lens, the image produced by the metalens for x – polarization can be described mathematically as a convolution between the optical intensity in the object plane and the quadratic PSF [19, 31]. Since the cylinders have almost the same dimension as the PSF, they appear blurred due to spatial frequency filtering. The object to image transformation, defined for the hyperbolic metalens in terms of a convolution with the normal incidence PSF only holds around the paraxial region because the PSF becomes highly distorted by the off-axis aberrations outside this region. Near the centre, the resolution of the metalens is around 550 nm, which leads to sharper images. Note that we can easily obtain different magnifications by adjusting the focal length of the two phase profiles independently, as detailed in section S3 of the SI.

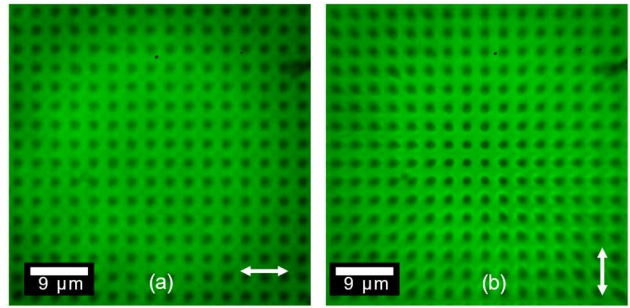

**Figure 3.** (a) (x-polarized light, quadratic profile) and (b) (y-polarized light, hyperbolic profile) show the c-Si cylinder array images obtained with the microscope setup using the metalens as an objective lens. The cylinders have diameters of 1 μm and are separated by 3 μm from each other.

**Point spread function**

Although Figs. 2 and 3 provide a qualitative assessment of the image properties, a more rigorous characterization requires the point spread function (PSF) to be measured. The PSFs measured for different angles of incidence is shown in Figure 4 (a) and (b) (see Methods for details of the optical setup) for both x-polarized (quadratic mode) and y-polarized (hyperbolic mode) incidences. In the quadratic mode, the metalens can maintain the PSF shape and full width at half maximum (FWHM) undistorted for angles as high as 40°. In this case, the PSF produces a FOV of 80º (the FOV is set by the NA, which is 0.75 here, see [16]). The FOV of the image, however, was limited to about 54º (the image is shown in Fig. S6 (a) of SI), mostly due to a decrease in the incoming power (see [16] for a fuller discussion). When operating as a hyperbolic metalens, the PSF is much smaller than in the quadratic mode, featuring a measured FWHM of 551 nm at normal incidence, which is half that of the quadratic lens (see Figure 4c). However, the PSF of the hyperbolic lens gets highly distorted for off-axis incidence (see Fig 4 b).

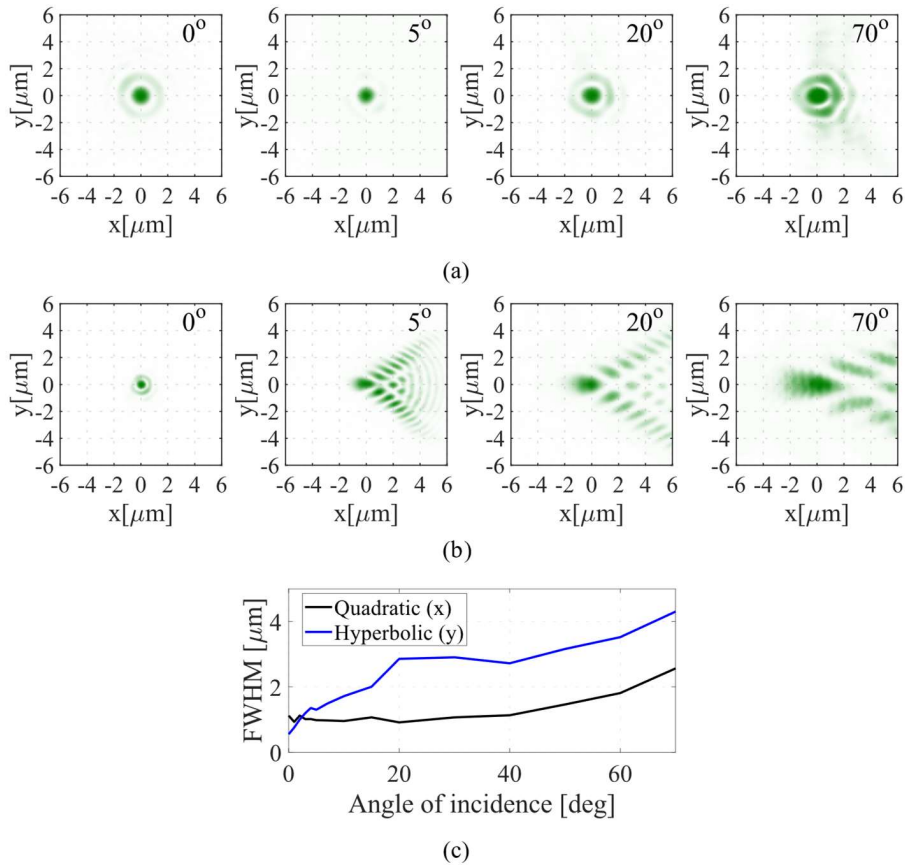

**Figure 4.** Measured point spread function (PSF) for different angles of incidence of the birrefringent metalens operating as (a) quadratic (x-polarization) and (b) hyperbolic (y-

polarization) metalens. (c) measured angular dependency of the longitudinal (x-axis in (a) and (b)) full width at half maximum of the birefringent metalens operating as a quadratic (black line) and hyperbolic (blue line) metalens.

**Conclusion**

In conclusion, we have demonstrated both wide field of view and high-resolution imaging with a single metalens by using polarization multiplexing operation at 532 nm wavelength. The metalenses use form-birefringent elliptical nanoposts to encode two separate phase profiles, i.e., a hyperbolic profile for high resolution and a quadratic profile for wide field of view, each accessed via one of the two orthogonal polarization states. When operating as a quadratic lens, we measured a resolution of approximately 1 μm and a FOV of 54°. When operating as a hyperbolic lens, we achieved near diffraction-limited resolution, with a PSF of FWHM=550 nm and the ability to resolve a 440 nm grating. The birefringent metalens can be used to build a compact and low-cost integrated microscope with two separate imaging modes, yet a single element metalens, a capability that may find applications in important fields such as endoscopy or integrated microscopy.

**Methods**

**Fabrication**

The metalenses were fabricated on commercially available 230 nm thick c-Si (100) wafers epitaxially grown on a sapphire substrate (from The Roditi International Corporation Limited.). The sample was cleaned with acetone, isopropyl alcohol (IPA) and oxygen plasma. It was subsequently spin-coated with a 300 nm positive electron beam resist layer (AR-P 6200.13, AllResist GmbH) followed by a 60 nm charge dissipation layer (AR-PC 5090, AllResist GmbH).

The structure was then patterned using an e-beam system (Voyager, Raith GmbH) followed by resist development in xylene. The pattern was transferred to silicon using reactive ion etching.

**Telescope imaging system**

The telescope setup we used to image far-field objects is shown in Figure S3. It consists of the metalens, a microscope stage, and a linear polarizer placed before the metalens. The object is positioned far away from the metalens ($d \gg f_m$, where $d$ is the object distance to the metalens, and $f_m$ is the metalens focal length), and its image is formed on the metalens focal plane with magnification $M_1 = -\frac{f_m}{d-f_m}$. Since the focal length is short, to avoid damage to the camera sensor, we built a microscope setup to relay the image formed by the metalens onto it. We emphasize that the relay stage is not a fundamental requirement, and it is used only for convenience. The relay stage magnification is given by $M_o = -\frac{f_T}{f_o}$, where $f_T$ is the tube lens focal length, and $f_o$ is the objective lens effective focal length. The total system magnification is then $M = -\frac{f_m}{d} M_o$. The images shown in Figure 2 were taken with $d = 7.3$ mm, $f_m = 660$ μm, which results in $M_1 \cong -\frac{1}{10.1}$ and $M_o = -60$ (we used a 60×, NA=0.85 Newport objective lens with a 160 mm tube lens). Therefore, the total system magnification is $M \cong 5.94$.

**Microscope imaging system**

The microscope setup was used to image small objects and it is shown in Figure S4. It consists of two stages: microscope 1, with the metalens used as an objective lens, and a Lens 1 (in practice, we used an objective lens system); and microscope 2, built with two spherical lenses (Lenses 2 and 3) to relay the image from microscope 1 to the sensor. Note the linear polarizer placed between Lens 2 and Lens 3. We placed Lens 1 very close to the metalens (focal length $f_m$) substrate side to collect light over a wide field of view. Lens 1 and Lens 2, which have focal lengths $f_{L1}$ and $f_{L2}$, respectively, are confocal. Finally, we placed the sensor on the focal plane of Lens 3, which has a focal length $f_{L3}$. Therefore, when an object is placed on the metalens focal plane, two real images are formed: one from microscope 1 and it is formed on Lens 1 back focal plane and the other one is formed on Lens 3 back focal plane, i.e., on the sensor. Because we used an objective lens system for Lens 1, which can collect light at wide angles due to its high NAs, we relay this image with microscope 2 to form a second image on the sensor. The magnification of the microscopes 1 and 2 are given by, respectively, $M_1 = -\frac{f_{L1}}{f_m}$ and $M_2 = -\frac{f_{L3}}{f_{L2}}$, which entails a total system magnification of $M = M_1 M_2 = \frac{f_{L1} f_{L3}}{f_m f_{L2}}$. The images shown in Figure 3 were taken with $f_m = 660$ μm and a 5×, NA=0.1, Newport objective lens, with an effective focal length of $f_{L1} = 25.4\ mm$, resulting in $M_1 = -38.48$. In the second stage, we used two plano-convex lenses with $f_{L2} = 15$ cm and $f_{L3} = 8.8$ cm, resulting in $M_2 = -0.58$. Therefore, the total system magnification is $M = 22.63$.

**Point Spread Function (PSF) Characterization Setup.**

The PSFs were measured using a microscope setup on a rotation stage, as shown in Figure S5. The metalens was illuminated with a laser at a wavelength of 532 nm, and the corresponding PSFs were then imaged onto the CMOS sensor. The laser polarization was controlled by a half waveplate.

## ASSOCIATED CONTENT

**Supporting Information.**

The following files are available free of charge.

Phase and transmission maps, Optical systems, Birefringent metalenses with different focal lengths (PDF).

AUTHOR INFORMATION

**Corresponding Author**

*Corresponding author: erm@usp.br

**Present Addresses**

$^\perp$Department of Physics, Harvard University, Cambridge, MA 02138, U.S.A.

**Author Contributions**

AM, TFK and ERM conceived the idea of the paper. JL, HL and BVB contributed in the analysis of results and design of the experiments. AM did all the calculations and most of the experiments. KL, DC and GA contributed to the fabrication and optical characterization. TFK and ERM supervised the project and wrote the manuscript.

ACKNOWLEDGMENT

This work is supported by São Paulo Research Foundation (FAPESP) (Grants 2013/07276-1, 2021/06121-0, 2015/21455-1, 2018/25372-1, 2020/00619-4 and 2020/15940-2); National Council for Scientific and Technological Development (CNPq) (304208/2021-3, 303562/2017-0). Guangdong Basic and Applied Basic Research Foundation (2020B1515020019). National

Natural Science Foundation of China (12074444). The authors D.C and T.F.K acknowledge financial support by the EPSRC of the UK (Grant EP/P030017/1 and P/T020008/1).

ABBREVIATIONS

FOV field of view; PSF point spread function; RCWA Rigorous Coupled Wave Analysis method; NA numerical aperture; FWHM full width at half maximum